\documentclass[twocolumn,reprint,aps,prd,nofootinbib,floatfix,superscriptaddress]{revtex4-1}
 \usepackage[unicode]{hyperref} 
\usepackage[utf8]{inputenc}
\usepackage{amsmath}
\usepackage{amsfonts}
\usepackage{amssymb}
\usepackage[left=2cm,right=2cm,top=2cm,bottom=2cm]{geometry}
\usepackage{braket}
\usepackage{dsfont}
\usepackage{hyperref}

\usepackage{graphicx}
\usepackage{tikz}
\usetikzlibrary{arrows,decorations.pathmorphing,backgrounds,positioning,fit,petri,shapes}
\usepackage{lipsum}

\usepackage{multirow}
\begin{document}

\title {Computing Nonlinear Power Spectra Across Dynamical Dark Energy Model Space with Neural ODEs}

\begin{abstract}
I show how to compute the nonlinear power spectrum across the entire $w(z)$ dynamical dark energy model space. Using synthetic $\Lambda$CDM data, I train a neural ordinary differential equation (ODE) to infer the evolution of the nonlinear matter power spectrum as a function of the background expansion and mean matter density across $\sim$$9 {\rm \ Gyr}$ of cosmic evolution. After training, the model generalises to {\it any} dynamical dark energy model parameterised by $w(z)$. With little optimisation, the neural ODE is accurate to within $4\%$ up to k = $5 \ h {\rm Mpc}^{-1}$. Unlike simulation rescaling methods, neural ODEs naturally extend to summary statistics beyond the power spectrum that are sensitive to the growth history.
\end{abstract}

\author{Peter L.~Taylor}
\email{taylor.4264@osu.edu}
\affiliation{Center for Cosmology and AstroParticle Physics (CCAPP), The Ohio State University, Columbus, OH
43210, USA}
\affiliation{Department of Physics, The Ohio State University, Columbus, OH 43210, USA}
\affiliation{Department of Astronomy, The Ohio State University, Columbus, OH 43210, USA}

\maketitle

\section{Introduction} \label{sec:intro}
Recent measurements from the Dark Energy Spectroscopic Instrument (DESI)~\cite{DESI:2024mwx, DESI:2025zgx}, supernovae~\cite{Scolnic:2021amr, Rubin:2023ovl, DES:2024jxu} and the Dark Energy Survey (DES)~\cite{DES:2025bxy} suggest the dark energy equation of state could be evolving with time. If true, measuring the evolution of $w(z)$ could help uncover the underlying physical theory (see~\cite{Copeland:2006wr} for a review).
\par Most dynamical dark energy measurements search for deviations from a cosmological constant using the Chevallier-Polarski-Linder (CPL) parameterisation (also called $w_{\rm 0}w_{\rm a}$CDM)~\cite{Chevallier:2000qy, Linder:2002et}. In this model, the dark energy equation of state (EOS) is given by $w(z) = w_{\rm 0} + w_{\rm a} / (1+z)$. If $(w_{\rm 0}, w_{\rm a}) \neq (-1, 0)$, this implies a deviation from the $\Lambda$CDM concordance model. However, if the EOS really is time-evolving, this simple parameterisation almost certainly does not describe the EOS, and could obscure the underlying physics.
\par A potentially more promising approach is to reconstruct the non-parametric redshift evolution of $w(z)$ directly as in~\cite{DESI:2025zgx, holsclaw2011nonparametric, DESI:2025fii, Zhao:2009ti, Zhao:2017cud}. It should however be noted that so far non-parametric reconstructions are in broad agreement with $w_{\rm 0} w_{\rm a}$CDM~\cite{DESI:2025zgx, DESI:2025fii}.  Alternatively, one can constrain parametric, but physically motivated, $w(z)$ models as in~\cite{DESI:2024kob,DESI:2025fii}. Using a combination of cosmic microwave background (CMB), baryonic acoustic oscillation (BAO) and supernovae (SNe Ia) measurements, these studies already constrain $w(z)$ to approximately $20 \%$ out to a redshift $z =1$. 
\par By extracting information from the growth of structure as well as the background expansion, upcoming photometric surveys including  Euclid~\cite{Euclid:2024yrr}, the Nancy Grace Roman Space Telescope\footnote{\url{https://roman.gsfc.nasa.gov/}} and the Rubin Observatory\footnote{\url{https://www.lsst.org/}} have the potential to substantially improve these constraints~\cite{Euclid:2019clj,Eifler:2020hoy,Xu:2022lne,LSSTDarkEnergyScience:2012kar}. To fully utilise the growth information, one must be able to model structure evolution for {\it all} $w(z)$ models deep into the nonlinear regime. 
\par In this paper I will focus on modeling the nonlinear matter power spectrum, but the techniques that I describe naturally extend to {\it any} summary describing the growth of structure.
\par One approach to compute nonlinear power spectra  is to train an emulator from high-resolution $N$-body simulations run across parameter space (see e.g.~\cite{Villaescusa-Navarro:2019bje, Euclid:2020rfv, Euclid:2018mlb, DeRose:2018xdj}). It would be incredibly computationally challenging to train emulators for all $w(z)$ models. For example,  {\tt EuclidEmulator2} is trained on 250 high-resolution simulations each taking 2000 node hours on GPU-accelerated nodes. 
\par Alternatively, one could follow a rescaling approach and rescale the $\Lambda$CDM power spectra following the {\tt ReACT} formalism~\cite{Cataneo:2018cic}, or rescale the particle-level simulation data following~\cite{Cataneo:2018cic}. The latter approach is used to train the {\tt BACCO} emulator where the particle-level rescaling is the primary accuracy limitation~\cite{Angulo:2020vky}.  {\tt BACCO} achieves percent level accuracy up to  $k_{\rm max} = 5 \ h {\rm Mpc} ^{-1}$ for $\Lambda$CDM and only just exceeds $5 \% $ for the CPL model\footnote{It should also be noted that the accuracy degrades as $w_{\rm 0}$ and $w_{\rm a}$ get further from the $\Lambda$CDM value and all the $w_{\rm 0}$ and $w_{\rm a}$ values considered in~\cite{Contreras:2020kbv} are closer to $\Lambda$CDM than the values preferred by the DESI DR2 BAO analysis.} (see Fig. 7 in~\cite{Contreras:2020kbv}).  
\par In is difficult, albeit not impossible, to extend {\tt ReACT} and particle rescaling formalism to the large number of higher-order and non-Gaussian statistics one would like to measure from photometric surveys. The {\tt ReACT} formlism would need to be adjusted on a statistic-by-statistic basis. Meanwhile, particle-rescaling requires access to 3D particle data, while for the more general (projected) summary statistics that are relevant for photometric surveys, it is  desirable to train on projected density shells (see e.g.~\cite{Fluri:2022rvb}).
\par  In this paper I propose a different approach that does not have these limitations. I argue that the redshift evolution of {\it any} summary statistic of the density field, $\boldsymbol{s}(z)$, satisfies a first order differential equation of the form
\begin{equation} \label{eqn:ode s}
\frac{d\boldsymbol{s}}{dz} = \boldsymbol{f}(\boldsymbol{s}, H(z), \bar \rho_m(z), z),
\end{equation} 
where $H(z)$ is the Hubble rate and $\bar \rho_m(z)$ is the mean matter density. Even though one cannot write this equation down analytically, one can approximate, $\boldsymbol{f}$, with a neural network. Numerically integrating this {\it neural ordinary differential equation} (ODE)~\cite{chen2018neural}, allows us to model the redshift evolution of the summary. To my knowledge this is only the second~\cite{Lanzieri:2022zvv} use of neural ODEs in cosmology. 
\par Crucially, $\boldsymbol{f}$, does not explicitly depend on cosmology, so we can train the network on  data generated by a $\Lambda$CDM emulator and, as I will show, the model generalises to {\it all} $w(z)$ cosmologies. Hence, to predict the value of $s(z)$ at any redshift in $w(z)$CDM, we compute the value of the summary at a high redshift, where $w(z)$CDM and $\Lambda$CDM are indistinguishable, before evolving the summary forward in time.
\par In this way, one can model {\it any} summary statistic for {\it all} $w(z)$. These in turn could be used to rapidly generate extremely realistic generative models~\cite{mousset2024generative} for simulation-based inference~\cite{Taylor:2019mgj, DES:2024xij}, but this is left for a future work. 
\par In this paper I focus on modeling the nonlinear power spectrum across $w(z)$ model space. The structure of the paper is as follows. In Sec.~\ref{sec:formalism}, I motivate the form of the ODE in Eqn.~\ref{eqn:ode s}, briefly review neural ODEs, and explain why the models generalise across dynamical dark energy model space when trained on $\Lambda$CDM data. The results are then presented in Sec.~\ref{sec:results}.

\section{Formalism} \label{sec:formalism}
\subsection{The Growth of Structure}
Consider the growth of structure in flat $w(z)$CDM at late times. In linear theory the $k$-modes are decoupled and the evolution of the matter density contrast for any $k$ satisfies the differential equation 
\begin{equation}  \label{eqn:linear}
\ddot{\delta}_k(t) + 2 H(t) \dot{\delta}_k(t) - 4\pi G \bar{\rho}_m(t) \, \delta_k(t) = 0,
\end{equation}
where the mean matter density is given by
\begin{equation} \label{eqn:mean matter}
\bar \rho_m(z) =  \frac{3 H_0^2}{8 \pi G} \Omega_{m} (1 + z)^3,
\end{equation}
and 
\begin{equation} \label{eqn:hubble}
\begin{aligned}
H^2(z) &= H_0^2 \Omega_m (1+z)^3 \\ &+ H_0^2\Omega_{\text{DE}} \exp\left(3 \int_0^z \frac{1 + w(z')}{1 + z'} \, dz'\right). 
\end{aligned}
\end{equation}
Here $H(z)$ is the Hubble rate, $\Omega_m$ is the matter density, and $\Omega_{\text{DE}}$ is the dark matter density. Crucially we see that at any redshift, $z$, the evolution of a linear mode, $\delta_k$, only depends on: the current state of $\delta$, the redshift, the Hubble rate and the mean matter density. {\it This is the key observation in this paper.}

\subsection{An ODE for the Evolution of Summary Statistics}
As we have seen, the growth of structure only depends on $z$, $\delta$, $\bar \rho_m(z)$ and $H(z)$. Hence, I make the ansatz that in $w(z)$CDM the evolution of a summary, $\boldsymbol{s}(z)$, satisfies an ODE of the form 
\begin{equation} \label{eqn:ode 1}
\frac{d \boldsymbol{s}(z)}{dz} = \boldsymbol{f} \big(\boldsymbol{s}(z), H(z), \bar \rho_m(z), z \big).
\end{equation} 
If we knew the forcing function $\boldsymbol{f}$, we could model the evolution of the summary by finding the summary at high-$z$, where it should match the $\Lambda$CDM prediction, before evolving the summary forward in time by numerically integrating the ODE. In practice, the function, $\boldsymbol{f}$, is not known and must be approximated by a neural network. This is described in the next section.

\subsection{Neural Ordinary Differential Equations}
Neural ODEs were introduced in~\cite{chen2018neural} to solve initial value problems (IVPs) of the form
\begin{equation}
\frac{d \boldsymbol{h}}{dt} = \boldsymbol{f} \big(\boldsymbol{h}(t),\boldsymbol{\theta}(t),t \big),
\end{equation}
where $\boldsymbol{h}$ is the state of the system at time $t$, h(0) is known and $\boldsymbol{\theta}(t)$ is a known function of $t$. One uses a neural network to learn the forcing function from training data, before using a numerical integrator to find the solution given by
\begin{equation}
\boldsymbol{h} (t_1) = \boldsymbol{h} (t_0) + \int_{t_0}^{t_1} \ {\rm d} t \ \boldsymbol{f}\big( \boldsymbol h(t), t, \boldsymbol{\theta} (t) \big),
\end{equation}
at later times. This is the exact formulation of the IVP in Eqn.~\ref{eqn:ode 1}, so this method can be used to learn the evolution of the summary from $\Lambda$CDM training data. In the next section I explain why we should expect this model to generalise to $w(z)$CDM.

\subsection{Generalisability of Neural ODEs Across Dynamical Dark Energy Model Space} \label{sec:generalise}
\par As I will demonstrate in Sec.~\ref{sec:w0wa}, the neural ODE accurately generalises to dynamical dark energy models, even when trained on $\Lambda$CDM data. This is the primary feature of the technique, so it is worth examining why the model is so flexible.
\par Recall the growth of structure is governed by two opposing forces: gravitational collapse and the background expansion. In linear theory (see Eqn.~\ref{eqn:linear}) these correspond to the Hubble friction and gravitational source terms. For every redshift, the neural ODE learns the mapping from the current state of the nonlinear power spectrum  and mean matter density (which together determine the gravitational source contribution) and the Hubble rate (which determines the Hubble friction contribution) to the instantaneous change in power. The model generalises because for any point in $w(z)$CDM parameter space there is a point in $\Lambda$CDM parameter space where these quantities (i.e. the power spectrum, mean matter density and Hubble rate) match, to a very good approximation. Looking at Eqns.~\ref{eqn:mean matter}-\ref{eqn:hubble} and noting the power spectrum is proportional to $\sigma_8$, suggests that these quantities can be matched to first order by rescaling $\sigma_8$, $\Omega_m$ and $H_0$. For example, at $z=1$, the mean matter density and Hubble parameter for a $\Lambda$CDM model with $\Omega_m=0.35$, $h = 0.65$ matches a $w$CDM model with $\Omega_m = 0.35$, $h = 0.65$ and $w=-1.1$, while the amplitude of the power spectrum can be made to match exactly with a simple rescaling of $\sigma_8$. The parameter correspondence changes with redshift, but this does not affect the accuracy of the model since the neural ODE models the instantaneous change in power, one redshift at a time.
\par This raises an important consideration when training emulators in $\Lambda$CDM. In order for neural ODEs to generalise across dynamical dark energy space, the power spectrum, mean matter density and Hubble rate for any $w(z)$ model of interest must lie inside the space spanned by the $\Lambda$CDM prior. For this reason it is worth training over broad priors in $\Lambda$CDM even if the parameters of interest are much better constrained in $\Lambda$CDM.\footnote{One may also need to clip the $\sigma_8$, $\Omega_m$ and $H_0$ prior in $w(z)$CDM to lie well within the $\Lambda$CDM prior to ensure the parameters correspondence is always maintained. This is left to a future work.} The neural ODE will also fail if $w(z)$ undergoes rapid oscillations, faster than the sampling rate in $z$, or if the equation of state is non-differentiable.

\section{Results} \label{sec:results}

\subsection{Training the Neural ODE to Infer the Evolution of the Nonlinear Power Spectrum}
From Eqn.~\ref{eqn:ode 1}, I write the evolution of the nonlinear power spectrum as 
\begin{equation} \label{eq:power ode}
\frac{ d \boldsymbol{P}(z) }{d z} = \boldsymbol{f} \Big( \boldsymbol{P}(z), H(z), \ln \big( \bar \rho_m(z) \Big), z \Big),
\end{equation}
where $\boldsymbol{P}(z)$ is a vector\footnote{Since the derivative is a function of the power spectrum across all $k$, this neural ODE naturally captures nonlinear mode-coupling.} in $k$ containing the values of $\ln \big( P(k,z) \big)$\footnote{I found training on the logarithm of the power spectrum, rather than the power spectrum, resulted in substantial improvements in the performance of the model.} for 30 logarithmically spaced $k$-bins in the range $k \in [0.01 \ h {\rm Mpc}^{-1}, 5 \ h {\rm Mpc}^{-1}]$ up to redshift $z=1.5$, deep into the matter-dominated epoch. Derivatives are computed using a two-sided finite difference with $\Delta z = 0.001$ at $20$ equally-spaced redshifts over this range.
\par To train the model, I generate $1.2 \times 10 ^5$  nonlinear power spectra in $\Lambda$CDM using the {\tt BACCO} emulator\footnote{https://baccoemu.readthedocs.io/en/latest/}, drawing cosmological parameters across the entire {\tt BACCO} prior (see Tab.~\ref{table:1}). This is treated as the ground truth for the remainder of this paper. I apply unit normal standardization to $H(z)$ and $\ln \Big( \bar \rho_m(z) \Big)$ across the training set before training 10 fully-connected networks in {\tt optax}\footnote{\url{https://github.com/google-deepmind/optax}}~\cite{deepmind2020jax} to approximate the forcing function, $\boldsymbol{f}$. The networks are five layers deep with 512 neurons per layer and ReLu activation functions. I use the {\tt Adam}~\cite{kingma2014adam} optimizer to perform gradient descent setting the learning rate to $10^{-3}$ with a batch size of 2048. I leave aside $10\%$ of the data for validation and apply an early stopping criterion of 20 epochs. All tensors are stored in 32-bit format and training takes approximately 10 minutes on an Nvidia RTX A6000.

\begin{table}[h]
\centering
\begin{tabular}{ll}
\hline
\textbf{Parameter} & \textbf{Range} \\
\hline
$\sigma_8$ & $[0.73,\ 0.9]$ \\
$\Omega_m$ & $[0.23,\ 0.4]$ \\
$\Omega_b$ & $[0.04,\ 0.06]$ \\
$n_s$ & $[0.92,\ 1.01]$ \\
$h$ & $[0.6,\ 0.8]$ \\
\hline
\end{tabular}
\caption{The {\tt BACCO} $\Lambda$CDM prior used in this work. I set $M_\nu = 0.0 \ {\rm eV}$. In $w_{\rm 0}w_{\rm a}$CDM, I restrict the analysis to the extended {\tt BACCO} prior: $w_{\rm 0} \in [-1.15, -0.85]$ and $w_{\rm a} \in [-0.3, -0.3]$.}
\label{table:1}
\end{table}

\par After training, I use {\tt diffrax}\footnote{https://docs.kidger.site/diffrax/}~\cite{kidger2021on} to numerically integrate the ODE using the {\tt Tsit5} integrator~\cite{tsitouras2011runge} with a time step of $10^{-2}$ before averaging over the models. It takes approximately 10 seconds to integrate all 10 models from $z=1.5$ to $z=0$ on an Nvidia RTX A6000. I now show how the model captures the evolution of the nonlinear power across the last $\sim$$9 \ {\rm Gyr}$ of structure evolution and generalises to dynamical dark energy models not seen during training. 

\subsection{Neural ODE Performance in $\Lambda$CDM}

\begin{figure}[!hbt]
\includegraphics[width = \linewidth]
{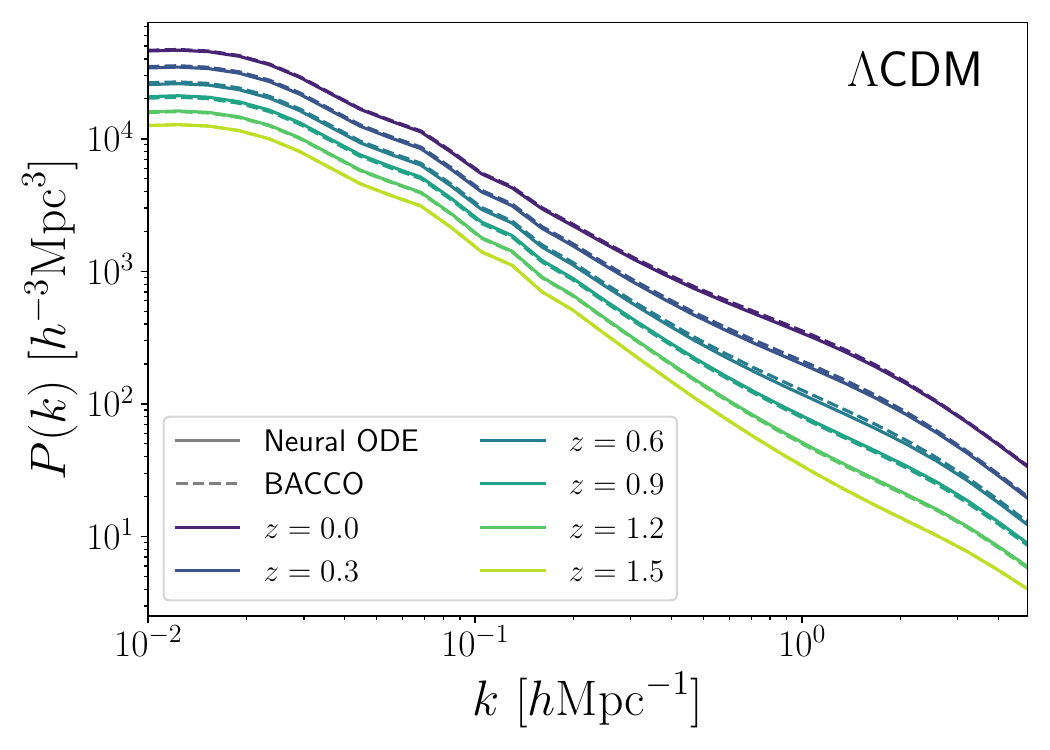}
\caption{{\bf Dashed lines:} {\tt Bacco} nonlinear matter power spectrum for a random cosmology, $(\Omega_m, \Omega_b, h_{\rm 0}, \sigma_8, n_{\rm s}) = (0.25, 0.04, 0.60, 0.79, 0.92)$, not seen during training. {\bf Solid Lines:} The neural ODE prediction. The model captures the evolution across $\sim$$9 \ {\rm Gyr}$ of structure evolution to within a few percent (see Fig.~\ref{fig:error}).}
\label{fig:pk}
\end{figure}

In Fig.~\ref{fig:pk}, I plot the evolution of the nonlinear power spectrum from z = 1.5 for a random $\Lambda$CDM cosmology not seen during training. The dashed lines correspond to the {\tt BACCO} power spectra, while the power spectra inferred from the neural ODE are plotted with solid lines. The model successfully tracks the evolution over $\sim 9 \ {\rm Gyr}$ of structure evolution, across half an order-of-magnitude change in power.

In the top panel of Fig.~\ref{fig:error}, I plot relative difference between the {\tt BACCO} spectra (which are treated as the ground truth in this paper) and the nonlinear ODE power spectra at $z=0$ for 10 random cosmologies not seen during training.  The residual error is less than $ 3 \%$ for all cosmologies. This is much smaller than baryonic feedback modeling uncertainties~\cite{Schneider:2018pfw,Huang:2018wpy} on small scales. Meanwhile on large scales, where the neural ODE exhibits $\sim$$1\%$ inaccuracies. In practice these can be entirely removed by using the exact linear theory on large scales and smoothly transitioning to the neural ODE prediction on small scales. 
\par The origin of the error is twofold. Even without the numerical inaccuracies introduced by the neural network approximation, numerical integration and finite difference numerical derivatives, one should not expect the neural ODE prediction to be exact. This is because the power spectrum is not a sufficient statistic to describe the current state of the system, and hence the growth of structure on nonlinear scales. For this reason, one would expect the accuracy to improve as more descriptive higher-order statistics are included.

\subsection{Neural ODE Performance in $w_{\rm0}w_{\rm a}$CDM} \label{sec:w0wa}
\begin{figure}[!hbt]
\includegraphics[width = \linewidth]
{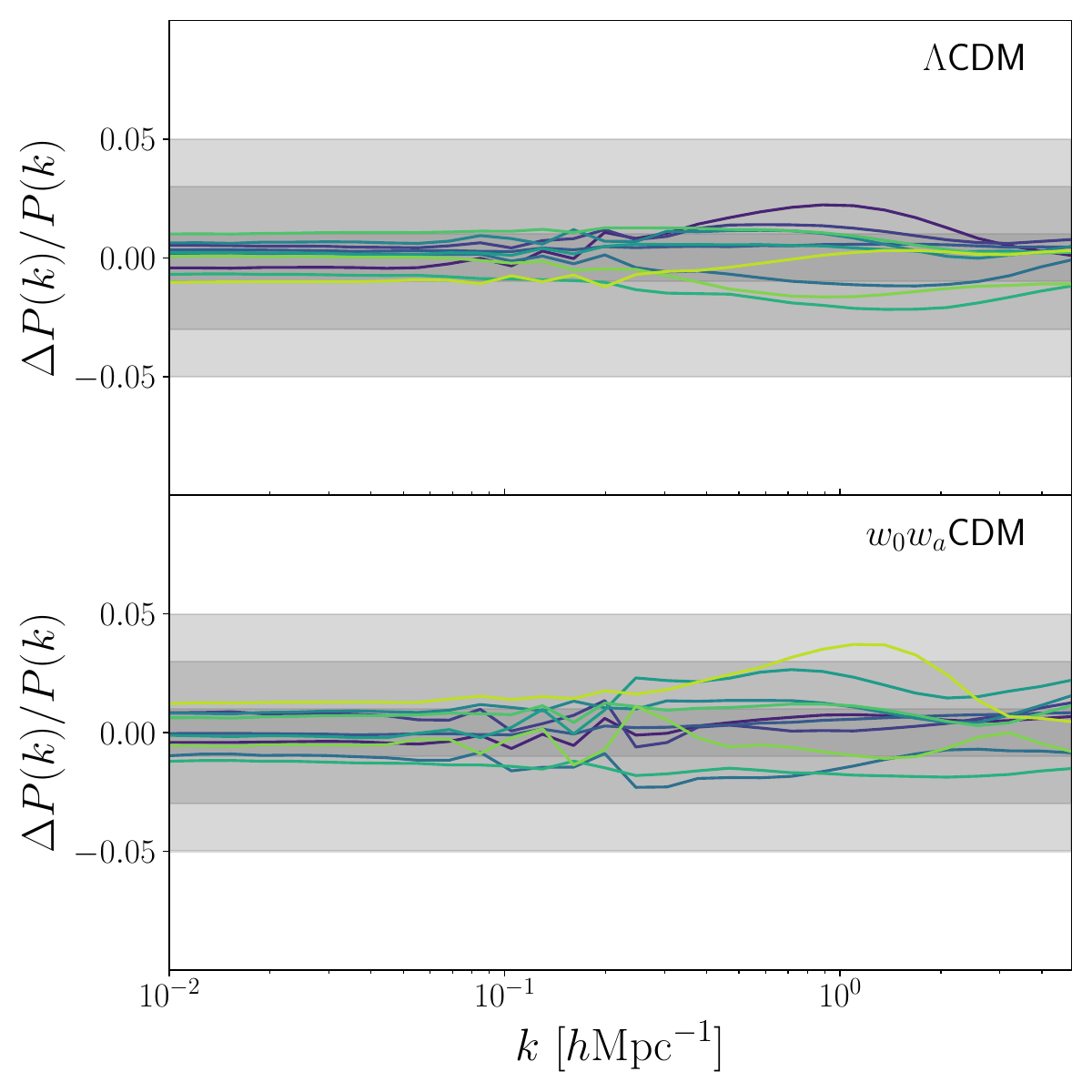}
\caption{{\bf Top Panel:} Residual power spectra errors for 10 random $\Lambda$CDM cosmologies not seen during training. The grey bands correspond to $1\%$, $3 \%$ and $5 \%$ errors. The Neural ODE is accurate to within $3 \%$ up to $k = 5 \ h{ Mpc}^{-1}$. {\bf Bottom Panel:} Same as above but for $w_{\rm 0} w_{a}$CDM. Despite being trained on $\Lambda$CDM data, the model generalises to dynamical dark energy modes (see Sec.~\ref{sec:generalise} for more details). The neural ODE is accurate to within $4 \%$ up to $5 \ h {\rm Mpc}^{-1}$.}
\label{fig:error}
\end{figure}

In the bottom panel of Fig.~\ref{fig:error}, I show the residual difference between {\tt BACCO} and the nonlinear ODE\footnote{When generating the neural ODEs predictions, I set the initial power spectrum at $z = 1.5$ to the $w_{\rm 0}w_{\rm a}$ initial conditions provided by {\tt BACCO} to avoid any residual bias not originating from the neural ODES inaccuracies. Ideally one would start from the $\Lambda$CDM at a higher redshift, but {\tt BACCO} is restricted to $z<1.5$. In the future work, I will switch to an emulator that does not have this restriction.} at $z=0$ for 10 random $w_{\rm 0} w_{\rm a}$CDM cosmologies drawn from the prior in Tab.~\ref{table:1}. Despite being trained on $\Lambda$CDM data, the model generalises to $w_{\rm 0} w_{\rm a}$CDM (see Sec.~\ref{sec:generalise}) and achieves $4 \%$ accuracy up to $k = 5 \ h {\rm Mpc}^{-1}$.

\subsection{Inferring the Power Spectrum in $w(z)$CDM}
I now show how to compute the power spectrum for a custom dark energy of state, $w_{\rm c} (z)$. 
\par
\begin{figure}[!hbt]
\includegraphics[width = \linewidth]
{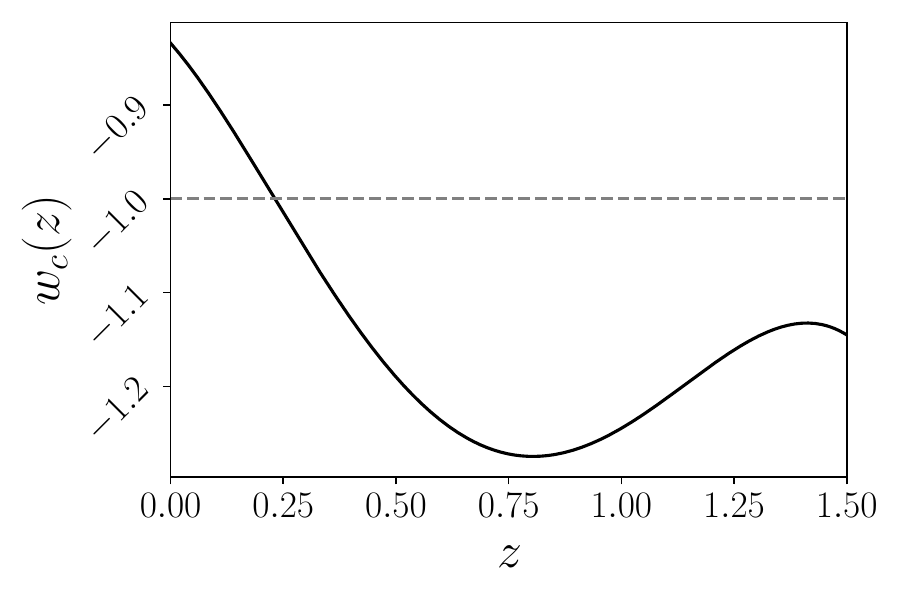}
\caption{A custom dark energy equation of state not described in the $w_{\rm 0}w_{\rm a}$ parameterisation. This $w(z)$ captures many of the qualitative features of the non-parametric reconstruction in the DESI DR2 BAO analysis (see Fig. 12 in~\cite{DESI:2025zgx}).}
\label{fig:wz}
\end{figure}

 In Fig.~\ref{fig:wz}, I plot an equation of state not described by the $w_{\rm 0}w_{\rm a }$ parameterisation. This $w(z)$ captures many of the key qualitative features currently preferred by the DESI DR2 analysis (see Fig. 12 in~\cite{DESI:2025zgx}), including phantom crossing below $z= 0.75$. This $w(z)$ is constructed by performing a fourth order least squares polynomial fit to the points $w = ( -0.8, -1., -1.1, -1.05, -1.2, -1.3, -1.4 -1.15, -1.15,$ $-1.15, -1.15)$ on an equally spaced grid in redshift up to $z=1.5.$ 
\par
\begin{figure}[!hbt]
\includegraphics[width = \linewidth]
{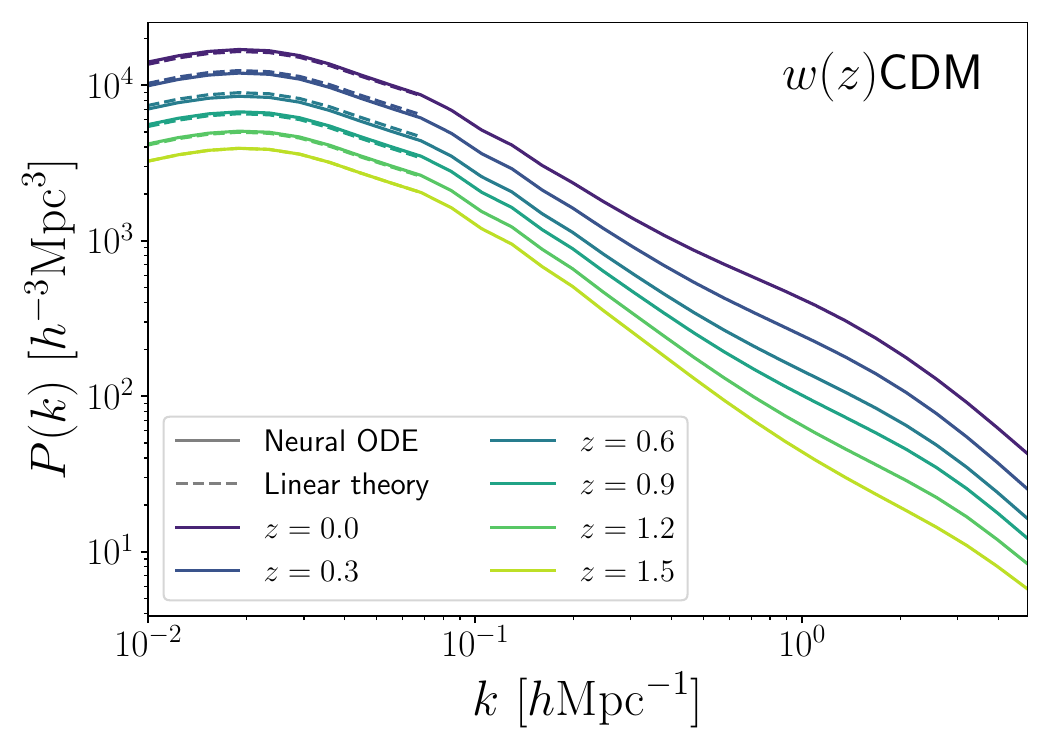}
\caption{The nonlinear power spectrum found using the neural ODE at a random point, $(\Omega_m, \Omega_b, h_{\rm 0}, \sigma_8, n_{\rm s}) = (0.36, 0.05, 0.77, 0.85, 0.93)$, in $w(z)$CDM space. I use the custom dark energy equation of state, $w_{\rm c} (z)$, which is plotted in Fig~\ref{fig:wz}. The dashed lines indicate the linear theory prediction down to scales of $k = 0.07 \ h {\rm Mpc}^{-1}$.}
\label{fig:pk_wz}
\end{figure}

 Since the value of $w_{\rm c}(z)$ lies in the range spanned by the $w_{\rm 0}$-$w_{\rm a}$ models tested in Sec.~\ref{sec:w0wa} for all $z$, it is safe to integrate the neural ODE to compute the nonlinear power spectrum. I show the resulting nonlinear power spectrum for a random point in parameter space in Fig.~\ref{fig:pk_wz}. To validate the model, the linear theory prediction down to scales of $0.07 \ h {\rm Mpc} ^{-1}$. As expected, the model is accurate to within $4 \%$ over the entire range.
\par 
\par In the future one could use this method to train emulators across $w(z)$ model space for a variety of non-parametric and physically motivated parametric models. This is left for a future work.

\section{Conclusion} \label{sec:conclusion}
\par I have shown how neural ODEs capture the redshift evolution of nonlinear matter power spectra. Inspired by linear theory, the key ansatz in this paper is that the instantaneous change in power is a function of the current state of the power spectrum and the mean matter density (which source the gravitational collapse), and the Hubble rate (which sets the strength of the Hubble friction).
\par Under this assumption, I describe the evolution of the power spectrum as the solution to a neural ODE initial value problem before showing how to train the model using $\Lambda$CDM data. Once trained, the model can be numerically integrated to compute the nonlinear power spectrum at any redshift and is accurate to within $3 \%$.
\par I have argued that this model generalises across dynamical dark energy models since for any point in $w(z)$CDM parameter space, there is a point in $\Lambda$CDM parameter space, where the power spectrum, mean matter density and Hubble rate match to a very good approximation. However for this to be true, we require training data over a broad prior, particularly $\sigma_8$, $H_0$ and $\Omega_m$. For this reason it is imperative that simulators continue to produce $\Lambda$CDM simulation suites that extend well beyond the regions of parameter space currently preferred by data.
\par I test the generalisability of the model explicitly in Sec.~\ref{sec:w0wa}, where I find that the neural ODE is accurate to $4 \%$ up to $k = 5 \ h {\rm Mpc} ^{-1}$ in $w_{\rm 0}w_{\rm a}$CDM. This is much smaller than current baryonic feedback modeling uncertainties on small scales. 
\par To showcase the efficacy of the technique, I use the neural ODE to compute the nonlinear power spectrum for a custom equation of state. In the future I will use this method to train emulators across dynamical dark energy model space. 
\par While I have focused on the nonlinear power spectrum, the method in principle extends to {\it any} summary statistic. Given the  plethora of $\Lambda$CDM training data, neural ODEs could be used to build dynamical dark energy emulators for many non-Gaussian summaries. These in turn could be  used to generate extremely realistic forward models~\cite{mousset2024generative} for simulation-based inference. This is left for a future work.
\par I have focused on dynamical dark energy but neural ODEs may prove adept at traversing more exotic model spaces. Ultimately neural ODEs may unlock nonlinear structure growth as a viable cosmological probe across a large landscape of models.

\section{Acknowledgments}
 PLT is supported in part by NASA ROSES 21-ATP21-0050. This work received support from the U.S. Department of Energy under contract number DE-SC0011726. PLT thanks Alex Jefferies for her typographical assistance and the anonymous referee whose comments have improved the paper.

 \bibliographystyle{apsrev4-1.bst}
\bibliography{bibtex.bib}

\appendix

\end{document}